\documentclass[a4paper]{jpconf}
\usepackage{graphicx,amsmath}

\newcommand\independent{\protect\mathpalette{\protect\independenT}{\perp}}
\def\independenT#1#2{\mathrel{\rlap{$#1#2$}\mkern2mu{#1#2}}}

\begin{document}
\title{Bell's theorem and the measurement problem: reducing two mysteries to one?}

\author{Eric G. Cavalcanti}

\address{Centre for Quantum Dynamics, Gold Coast campus, Griffith University, Southport, Queensland 4215, Australia}

\ead{e.cavalcanti@griffith.edu.au}

\begin{abstract}
In light of a recent reformulation of Bell's theorem from causal principles by Howard Wiseman and the author, I argue that the conflict between quantum theory and relativity brought up by Bell's work can be softened by a revision of our classical notions of causation. I review some recent proposals for a quantum theory of causation that make great strides towards that end, but highlight a property that is shared by all those theories that would not have satisfied Bell's realist inclinations. They require (implicitly or explicitly) agent-centric notions such as ``controllables'' and ``uncontrollables'', or ``observed'' and ``unobserved''. Thus they relieve the tensions around Bell's theorem by highlighting an issue more often associated with another deep conceptual issue in quantum theory: the measurement problem. Rather than rejecting those terms, however, I argue that we should understand why they seem to be, at least at face-value, needed in order to reach compatibility between quantum theory and relativity. This seems to suggest that causation, and thus causal structure, are emergent phenomena, and lends support to the idea that a resolution of the conflict between quantum theory and relativity necessitates a solution to the measurement problem.
\end{abstract}

\section{Introduction}

\begin{quotation}``For me then this is the real problem with quantum theory: the apparently essential conflict between any sharp formulation [of quantum theory] and fundamental relativity. That is to say, we have an apparent incompatibility, at the deepest level, between the two fundamental pillars of contemporary theory (...) It may be that a real synthesis of quantum and relativity theories requires not just technical developments but radical conceptual renewal.''- J.S. Bell, 1986 \cite{Bell1986}
\end{quotation}

Bell's famous 1964 theorem \cite{Bell1964} has challenged our understanding of quantum theory and ignited much debate on its foundations as well as technological spin-offs such as device-independent quantum cryptography and randomness amplification. More than 50 years later, the debate still hasn't settled. As the quote above shows, Bell saw his work as demonstrating a fundamental incompatibility between quantum theory and relativity, and believed that a solution would require a ``radical conceptual renewal''. Here we take the point of view that at least part of that renewal regards our understanding of causality. In fact, Bell's theorem is best understood as being about the constraints from causal structure to possible correlations among events, where relativity comes in merely to supply the causal structure.

But who cares about causality? Some philosophers, like Bertrand Russell, have argued that it has no role in fundamental physics. In his famous 1913 essay ``On the Notion of Cause'', Russell that ``The law of causality, I believe, like much that passes muster among philosophers, is a relic of a bygone age, surviving, like the monarchy, only because it is erroneously supposed to do no harm" \cite{Russell1912}. For Russell, the time-asymmetric notions of cause and effect are not compatible with the deterministic and time-symmetric laws of physics, in which the future determines the past just as the past determines the future. Another point of view is defended by Nancy Cartwright \cite{Cartwright1979}: ``Bertrand Russell argued that laws of association are all the laws there are, and that causal principles cannot be derived from the causally symmetric laws of association. (...) Causal principles cannot be reduced to laws of association; but they cannot be done away with".  For Cartwright, causal principles are required to distinguish between {\em effective} and {\em ineffective} strategies.

For Cartwright, effective action requires that agents represent their world in causal terms. This view is shared by philosophers such as Huw Price, although he argues \cite{Price2007} that nevertheless this position does not require realism about causation, that is, it doesn't require a commitment to causation and its associated agent-centric concepts as a fundamental aspect of the ontology. Rather, Price defends the view that causation emerges from an agent's perspective in the world, while proposing that the world itself may be describable in a Òview from nowhereÓ, a block universe.

Despite the influence of Russell's causal eliminativist view in the philosophical literature, I believe that it is a minority view within physicists. The word ``causality'' certainly appears in a wide range of physics literature, often associated with the light-cone structure of relativity theory and the impossibility of {\it something} (information, physical systems, influences...) being transmitted between space-like separated regions. Specifically within discussions on Bell's theorem, the debate is usually about {\em what notion of causality} to use, rather than whether causality plays a role in Physics. The views in this realm can be broadly classified according to the holder's philosophical inclination as an {\it operationalist} or a {\it realist}. Here the conflict becomes a terminological dispute about what is the physically correct notion of {\it locality}, with different camps disagreeing about whether or not it is violated by quantum theory. Roughly speaking, the realist (like Bell) focuses on the importance of causal notions as the basis for the explanation of correlations, and believes that quantum correlations violate the causal structure imposed by relativity and is thus nonlocal. The operationalist, on the other hand, focuses on the fact that quantum correlations cannot be used to send signals faster than light and is thus local, and believes that it is Bell's classical notion of {\it local causality} that is to blame for the apparent conflict.

In a recent article \cite{Causarum}, Howard Wiseman and the author have presented a reformulation of Bell's theorem in terms of causal notions, and proposed a way forward for the debate between operationalists and realists. In that formulation, Bell's theorem was shown to be derived from four Axioms and four Principles the definitions of which can be agreed upon by operationalists and realists alike, with the disagreement being merely about which of those to drop in light of the contradiction with observed phenomena. For the realist, it is relativistic causality that must go, to be supplemented perhaps by a preferred foliation of space-time such as that required by Bohmian mechanics. For the operationalist, it is the notion that a cause explains the correlation between two variables by rendering them uncorrelated. In our reformulation, this was accomplished by following an earlier proposal by the author and Ray Lal to break {\it Reichenbach's principle of common cause} into two logically separate principles \cite{CavalcantiLal}.

Reichenbach's principle \cite{Reichenbach1956} is a fundamental tenet in the classical theory of causation. It states that if two events are correlated, then either one causes the other, or there exists a common cause such that conditioned on it, the events become uncorrelated. In \cite{Causarum} this principle was separated as a principle of {\it Common Causes}, requiring that two sets of events $\mathcal{A}$ and $\mathcal{B}$ are correlated only if one causes the other or there exists a set of common causes $\mathcal{C}$ that {\it explains} the correlation (and where the term {\it explains} is left to be defined by further principles); and a principle of {\it Decorrelating Explanation}, by which a set of causes $\mathcal{C}$ explains the correlation between $\mathcal{A}$ and $\mathcal{B}$ only if those variables are left uncorrelated by conditioning on $\mathcal{C}$. By dropping the requirement of Decorrelating Explanation, the operationalist can thus keep relativistic causal structure.

But Reichenbach's principle is at the basis of the application of causal reasoning in the sciences. It is for example, a consequence of the basic assumptions behind the theory of causal networks (see e.g. the book of Pearl \cite{Pearl}). That theory has a wide range of applicability, and simply dropping Reichenbach's principle, even if only partly, would be ``throwing out the baby with the bathwater''. The challenge then becomes to generalise the classical theory of causal networks to accommodate quantum mechanics, in a way that reduces to the classical case in some appropriate limit. Recent work in quantum causal structures has followed that line of reasoning and proposed generalisations of the classical theory of causal networks into a quantum theory \cite{Fritz2014,Henson2014,Pienaar2015}.

These important developments go a long way towards alleviating the ``essential conflict'' raised by Bell, and at this stage the operationalist would likely be satisfied that a resolution has been achieved. Here I will argue however, that there is still cause for concern. The reformulations so far seem to fall short of another of Bell's concerns, that the operationalist would do well not to ignore. That is, they seem to require an apparent agent-centric fundamental distinction (at least implicitly) between ``controllable'' and ``uncontrollable'' events, and indeed causal concepts are only applicable at the level of ``measurements''. This suggests that the resolution of the mystery of Bell's theorem so far has been a reduction to another foundational puzzle in quantum theory -- the measurement problem. This raises the question of whether this is a necessary condition for the application of causal concepts that are compatible with quantum physics, and if so, does it point to causal concepts having an emergent nature?

This paper is structured as follows: In Sec.~\ref{sec:classical} I review the classical theory of causal networks and reformulate Bell's theorem within that language. In Sec.~\ref{sec:quantum} I look at recent proposals for quantum theories of causal networks and argue that although they provide a reformulation of the concept of causal explanation that allows for compatibility with relativistic causal structure and reduces to the classical case in the appropriate limit, they all require agent-centric causal concepts. I end with a discussion about how this reduces the puzzle of Bell's theorem to the measurement problem, and points to the possibility that causal concepts are emergent.

\section{Classical causal networks} \label{sec:classical}

\begin{quotation}
``Do we then have to fall back on ``no signalling faster than light'' as the expression of the fundamental causal structure of contemporary theoretical physics? That is hard for me to accept. For one thing we have lost the idea that correlations can be explained, or at least this idea awaits reformulation''. Ð J.S. Bell, ``La Nouvelle Cuisine'', in \cite{Bell1987}.
\end{quotation}

What does it mean to explain correlations? Suppose a rare virus infection spreads almost simultaneously in Australia and Brazil. Possible explanations epidemiologists would entertain include: a) The virus was carried from Australia to Brazil; b) The virus was carried from Brazil to Australia or c) The virus was carried to both places from a third country. This idea is encapsulated by Reichenbach's Principle of Common Cause (RPCC) \cite{Reichenbach1956}. Informally, it says that if two events are correlated, either one is a direct cause of the other, or they share a common cause. This principle is fundamental in many areas of science and underlies the important idea that correlation does not necessarily imply causation.

\begin{figure}[h]
\includegraphics[width=6cm]{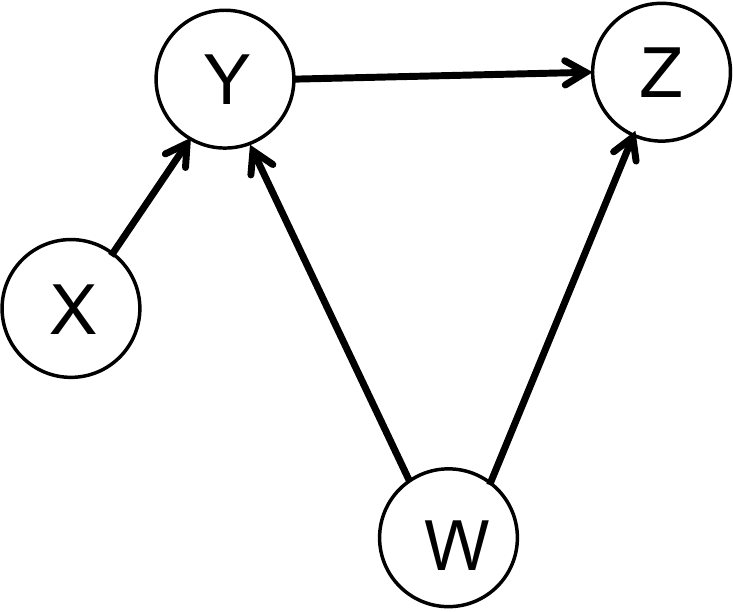}\centering
\caption{\small A directed acyclic graph (DAG). Nodes represent random variables associated with possible events and directed edges represent causal connections between events.}
\label{DAG}
\end{figure}

A more modern and general theory of causation and its role in explaining correlations can be found in the theory of causal networks \cite{Pearl}. It has been developed as a tool to connect causal inferences and probabilistic observations. For example, in developing a drug, a pharmaceutical company may want to distinguish, from some observed statistics, the effects of the drug from correlations that do not imply causation (e.g. because a subset of the population that is more likely to be cured without the drug is also more likely to use the drug in the first place). Here causal structure is encoded as a graph, with nodes representing random variables of interest and directed edges between nodes representing causal links. Since we want to exclude causal loops, the resulting structure is that of a \textit{directed acyclic graph} (DAG). An example of a DAG is given in Fig.~\ref{DAG}.

In a DAG $G$, the nodes that have arrows pointing to a given node $X$ are denoted the \textit{parents} of $X$. The set of all parents of parents (of parents, etc) of $X$ is the set of \textit{ancestors} of $X$. The \textit{descendants} of $X$ are all variables for which $X$ is an ancestor.

A DAG can be seen as representing the conditional independences associated with any probability distribution $P(\cdot)$ over the variables in the graph that is compatible with that causal structure. We will use lower case letters to denote particular values of a random variable, e.g. $X=x$. Formally, we say that two sets of variables $X$ and $Y$ are \textit{conditionally independent} in a probability distribution $P(\cdot)$ given a set of variables $Z$ if and only if knowledge of $Y$ provides no new information about $X$ given that we already know $Z$, that is $P(x|y,z)=P(x|z)$. This is represented as $(X \independent Y|Z)_P$.

Conditional independences satisfy the \textit{graphoid axioms}:

\vspace{.3cm}

Symmetry: $(X \independent Y|Z)_P \iff (Y \independent X|Z)_P$;

Decomposition: $(X \independent YW|Z)_P \implies (X \independent Y|Z)_P$;

Weak union: $(X \independent YW|Z)_P \implies (X \independent Y|ZW)_P$;

Contraction: $(X \independent Y|ZW)_P \; \& \; (X \independent W|Z)_P \implies (X \independent YW|Z)_P$;

Intersection: $(X \independent W|ZY)_P \; \& \; (X \independent Y|ZW)_P \implies (X \independent YW|Z)_P$

\vspace{.3cm}

A DAG represents the conditional independences of a probability distribution in the following way. Every probability distribution $P$ over n variables $X_1,X_2,...X_n$ can be decomposed as a product of conditional distributions:

\begin{equation}\label{pfact}
P(x_1,...,x_n) = \prod_j P(x_j|x_1,...x_{j-1})\,.
\end{equation}

Some of those conditional distributions will display independences, say for example $P(x_3|x_1,x_2)=P(x_3|x_1)$. The conditional distribution for a variable $X_j$ will depend only on a subset of variables, the parents of $X_j$, denoted by $PA_j$. A DAG $G$ represents $P$, or is compatible with $P$ if and only if $P$ admits a factorisation of the following form, where the parents $PA_j$ are defined by the variables in $G$ with directed edges terminating at $X_j$:

\begin{equation}\label{GrepresentsP}
P(x_1,...,x_n) = \prod_j P(x_j|pa_j)\,.
\end{equation}

In classical causal models, one obtains the conditional independencies from the graph using a rule called \textit{d-separation}. Two sets of variables $X$ and $Y$ are d-separated given a set of variables $Z$ (denoted by $(X\independent Y|Z)_d$) if and only if $Z$ ``blocks'' all paths (sequences of consecutive edges) $p$ from $X$ to $Y$. A path $p$ is said to be d-separated by $Z$ if and only if (i) the path $p$ contains a chain $i \rightarrow m \rightarrow j$ or a fork $i \leftarrow m \rightarrow j$ such that the middle node $m$ is in $Z$, or (ii) the path $p$ contains an inverted fork (head-to-head) $i \rightarrow m \leftarrow j$ such that the node $m$ is not in $Z$, and there is no directed path from $m$ to any member of $Z$.

The implication of d-separation for conditional independences is that if $X$ and $Y$ are d-separated by $Z$ in a DAG $G$, then for all distributions $P$ that are compatible with $G$, $X$ is independent of $Y$ given $Z$. That is, $(X\independent Y|Z)_d \implies (X\independent Y|Z)_P$ for all $P$ compatible with $G$. In other words, d-separation is a sound criterion for conditional independence. Furthermore, if for all distributions $P$ compatible with $G$, $(X\independent Y|Z)_P$, then $(X\independent Y|Z)_d$. In other words, d-separation is a complete criterion for conditional independence.

The conditional independencies obtained from d-separation satisfy the following property:

\textbf{Causal Markov Condition}: In any probability distribution $P$ that is compatible with a graph $G$, A variable $X$ is independent of all its non-descendants, conditional on its parents.

The causal Markov Condition implies a weaker property that we call \textit{Causal Completeness}:

\textbf{Causal Completeness}: A variable $X$ is independent of its non-descendants, conditional on its ancestors.

The reason for considering this weaker notion is that in theories of causal networks beyond the classical case, it may not always be convenient to assume that the parent nodes screen off causal influences from more distant ancestors (see e.g. \cite{Pienaar2015}). With the interpretation that the parents of a variable $X$ are its direct causes, and the ancestors are all of its possible causes, this principle states that an event is independent of its non-effects given its causes. Causal Completeness implies Reichenbach's Principle of Common Cause, with the common causal past of two events $X$ and $Y$ interpreted as the intersection of their respective sets of ancestors.

\subsection{Causal networks and Bell's theorem}

\begin{figure}[h]
\includegraphics[width=6cm]{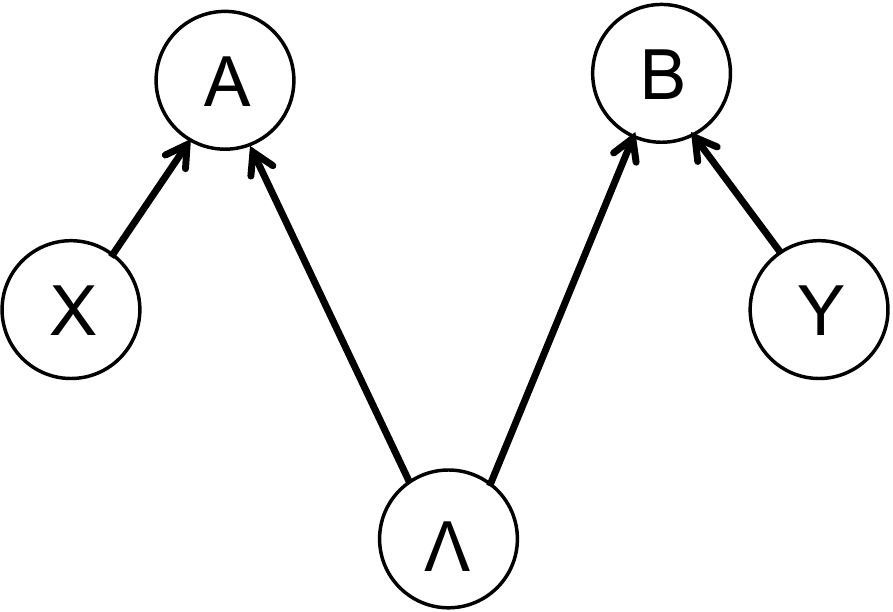}\centering
\caption{\small A directed acyclic graph (DAG) representing the causal structure of a Bell correlation scenario, under the assumption of relativistic causality.}
\label{fig:Bell}
\end{figure}

We are now ready to formulate Bell's theorem in this language. Consider the usual Bell scenario, with Alice and Bob in two space-like separated regions, performing experiments on two entangled quantum systems. Their respective choices of experiment are given by variables X and Y, with the associated outcomes represented by A and B. Assuming relativistic causal structure, there cannot be any direct causal connection between events at Alice's lab and events at Bob's lab. There could, however, be some common causes $\Lambda$ in their common past light cone (for example, variables associated with the preparation of the entangled quantum state). We assume that X and Y are free variables, meaning that there are no arrows going into the causal graph representing the situation. Given these assumptions, the scenario is represented by the graph in Fig.~\ref{fig:Bell}.

The above considerations allow us to determine the set of probability distributions $P$ that are compatible with this graph. We can always decompose the distribution over the random variables in question as 
\begin{equation}
P(a,b,x,y,\lambda) = P(a|b,x,y,\lambda)P(b|x,y,\lambda)P(\lambda|x,y)P(x|y)P(y)\,.
\end{equation}

Causal Completeness implies that $P(a|b,x,y,\lambda)=P(a|x,\lambda)$, $P(b|x,y,\lambda)=P(b|y,\lambda)$, $P(\lambda|x,y)=P(\lambda)$ and $P(x|y)=P(x)$, and so
\begin{equation}
P(a,b,x,y,\lambda) = P(a|x,\lambda)P(b|y,\lambda)P(\lambda)P(x)P(y) \,.
\end{equation}
Averaging the conditional probabilities $P(a,b,\lambda|x,y)=P(a,b,x,y,\lambda)/P(x,y)$ over $\lambda$ we thus obtain the usual factorisability condition of a Local Hidden Variable model:

\begin{equation}
P(a,b|x,y) = \sum_\lambda P(\lambda)P(a|x,\lambda)P(b|y,\lambda)\,,
\end{equation} 
and as is well known, this leads to the Bell inequalities, which are violated by certain quantum correlations. Therefore, assuming relativistic causal structure and free choices, quantum correlations cannot be reproduced by the classical theory of causation, which assumes Causal Completeness.

\section{Quantum causal networks}\label{sec:quantum}

An alternative to resolve the conflict while maintaining relativistic causality is to reject some of the principles underlying the classical theory of causal networks, including Causal Completeness. Several recent works have proposed generalisations of the theory of causal networks to accommodate quantum correlations \cite{Fritz2014, Henson2014, Pienaar2015}. In the following I will argue that they raise an interesting puzzle -- the apparent need for agent-centric notions such as controllable/uncontrollable or observed/unobserved in quantum theories of causal structure.

Pienaar and Brukner (PB) \cite{Pienaar2015} developed a theory based on a DAG with two different types of nodes, corresponding to events that are controllable ``settings'' and those that are observable but not directly controllable ``outcomes''. Based on this they proposed a theory of quantum causal models that satisfies a:

\textbf{Quantum Causality Condition}: An outcome is independent (conditional on the empty set) of all settings that are not its causes and all outcomes that do not share a common cause.

Note that unlike the case of Causal Completeness, here a fundamental distinction is made between settings and outcomes. Underlying this condition there is an associated notion of `q-separation'  that generalises the notion of d-separation:

\begin{quotation}
Given a DAG representing a quantum network, two disjoint sets of variables $X$ and $Y$ are said to be {\it q-separated} by a third disjoint set $Z$, denoted $(X\independent Y|Z)_q$, if and only if every undirected path between $X$ and $Y$ is rendered inactive by a member of $Z$. A path connecting two variables is rendered inactive by $Z$ if and only if at least one of the following conditions is met:

(i) both variables are settings, and at least one of the settings has no directed path to any outcome in $Z$;

(ii) one variable is a setting and the other is an outcome, and there is no directed path from the setting to the outcome, or to any outcome in $Z$;

(iii) the path contains a collider $i \rightarrow m \leftarrow j$ where $m$ is not an outcome in $Z$, and there is no directed path from $m$ to any outcome in $Z$.
\end{quotation}

The authors then proceed to show that q-separation is a sound and complete criterion for conditional independences within their notion of quantum causal models. 

Another framework is given by Henson, Lal and Pusey (HLP)~\cite{Henson2014}, who keep d-separation, but at the expense of introducing another form of agent-centric concepts: ``observed'' and ``unobserved'' nodes. Classical data, such as outcomes of measurements, are associated to observed nodes. Unobserved nodes output only systems, and have no outcomes associated to them. HLP define a notion of a distribution $P$ being {\it generalized Markov} with respect to a graph $G$, with reference to this distinction. They also show that their theory reduces to the classical theory  if all nodes are observed, in which case a generalized Bayesian network is a classical Bayesian network.

Here we note that the work of Fritz \cite{Fritz2014} provides a framework that does not make a distinction between different types of nodes. However, no discussion is given there for a substitute for d-separation or the Causal Markov Condition, and it is not clear therefore whether that framework can accommodate a notion of quantum causality that is sounds and complete for the conditional independences implied by quantum phenomena while not making agent-centric distinctions such as the two frameworks discussed above.

For another discussion of causal concepts underlying Bell's theorem and further motivation for the idea that agent-centric notions are required to maintain compatibility with relativistic causality is given in a recent paper of Wiseman and the author \cite{Causarum}.

\section{Who do we think ``we'' are?}

\begin{quotation}
``More importantly, the ``no signalling...'' notion rests on concepts which are desperately vague, or vaguely applicable. The assertion that ``we cannot signal faster than light'' immediately provokes the question: Who do we think {\it we} are? {\it We} who can make ÔmeasurementsÕ,  {\it we} who can manipulate Ôexternal fieldsÕ,  {\it we} who can ÔsignalÕ at all, even if not faster than light? Do {\it we} include chemists, or only physicists, plants, or only animals, pocket calculators, or only mainframe computers?'' - J.S.~Bell~, ``La Nouvelle Cuisine'', in \cite{Bell1987}.
\end{quotation}

The theories of quantum causal networks discussed in the previous section are important developments in the elucidation of Bell's theorem, and arguably provide at least a partial answer to the question of what counts as a causal explanation of correlations in the quantum case, an idea that ``awaited reformulation'' according to Bell, as quoted in the beginning of Sec.~\ref{sec:classical}. 

However, the frameworks above seem to indicate that we can recover causal principles that conciliate quantum mechanics and relativistic causality, but at the expense of introducing agent-centric notions of causation. For an operationalist, this doesn't seem to be a very high price to pay. However, I would like to suggest that there's cause for further consideration. 

As the quote above shows, Bell had two concerns regarding the usage of ``no signalling'' as ``the expression of fundamental causal structure of contemporary theoretical physics''. One concern, addressed by quantum causal networks, is the reformulation of the idea of causal explanation, generalising Reichenbach's principle of common cause, which was implicit in Bell's thought. The other concern, as the passage above shows, is that the concept of ``no signalling'' relies on agent-centric notions, and therefore should not form part of a fundamental physical theory, no more than the concept of ``measurement''.

Again, this may be an arcane concern for an operationalist. But remember, that if the operationalist had never paid attention to the realist point of view, they would never have realised that there was a whole theory of causation that needed to be modified to accommodate quantum theory.

Now I don't entirely agree with Bell either, mind you. Clearly we {\it can} talk about ``settings'' and ``outcomes'' rather unambiguously, and it's useful to do so. This of course, does not require anthropocentrism -- computers and measurements devices are good enough ``agents'' for defining what we mean by those terms, without the need to involve human beings. But this might be hinting, at least to my mind, that a complete conciliation between relativity and quantum theory will require a deeper understanding of the role these agent-centric notions play in the theory. And here again, I would urge the operationalist to heed the lessons learned from the realist. Realists are not willing to accept agent-centric notions as fundamental concepts in a physical theory. If these causal concepts are stubbornly agent-centric, for a realist that means that they ought to be emergent properties. If that's the case, we must understand how and why they emerge. It's a speculation, but I think it's one worth considering. 

Maybe one way of seeing all this is that these new developments have gone to great lengths towards solving the mystery of Bell's theorem, towards the conceptual renewal that Bell sought in order to reconcile relativistic causal structure and quantum theory. However, the efforts so far have done so only at the expense of forcing us to face up to the other major conceptual puzzle of quantum theory: the measurement problem. Which is a great advance: perhaps we are reducing two mysteries to one!

Going further, it would be interesting to determine whether Causal Completeness can be maintained together with relativistic causal structure, perhaps with a modification of the concept of independence, or relaxing some of the implicit or explicit assumptions of the formalism, e.g.~that the events considered are part of a single relativistic space-time manifold -- would a multiverse theory allow for fundamental causal explanation free of agent-centric terms? Or perhaps it could provide a story about how those notions emerge from a more fundamental background?

{\it Acknowledgement.} The author would like to thank Howard Wiseman, Sally Shrapnel, Jacques Pienaar, Fabio Costa and Magdalena Zych for useful discussions. This publication was made possible through the support of a grant (FQXi-RFP-1504) from the Foundational Questions Institute Fund (fqxi.org) at the Silicon Valley Community Foundation, and Australian Research Council projects DE120100559 and DP140100648.

\end{document}